\documentclass{article}
\usepackage{spconf,amsmath,graphicx,amssymb,pgfplots, xcolor}
\pgfplotsset{width=8.4cm, height=5cm, compat=1.9}

\title{AGE-CONDITIONED SYNTHESIS OF PEDIATRIC COMPUTED TOMOGRAPHY WITH AUXILIARY CLASSIFIER GENERATIVE ADVERSARIAL NETWORKS}
\name{Chi Nok Enoch Kan, Najibakram Maheenaboobacker, Dong Hye Ye\thanks{This study was funded by the NIH grant 1U01EB023822.}}

\address{Department of Electrical and Computer Engineering, Marquette University, Milwaukee, WI 53233}

\begin{document}
%\ninept
%
\maketitle
\begin{abstract}
Deep learning is a popular and powerful tool in computed tomography (CT) image processing such as organ segmentation, but its requirement of large training datasets remains a challenge. Even though there is a large anatomical variability for children during their growth, the training datasets for pediatric CT scans are especially hard to obtain due to risks of radiation to children. In this paper, we propose a method to conditionally synthesize realistic pediatric CT images using a new auxiliary classifier generative adversarial network (ACGAN) architecture by taking age information into account. The proposed network generated age-conditioned high-resolution CT images to enrich pediatric training datasets.
\end{abstract}
%
% \begin{keywords}
% One, two, three, four, five
% \end{keywords}
%
\section{Introduction}
\label{sec:intro}

Deep learning shows promising results in CT image
processing tasks such as organ segmentation~\cite{C1}. However, one major challenge in applying deep learning to CT images is the lack of training data particularly for pediatric patients. Even though pediatric CT organ segmentation methods generally require large amounts of training data to deal with the anatomical variability over children's growth, pediatric CT scans are hard to obtain as children are more vulnerable to ionizing radiation exposure in CT scans than adults \cite{C2}. Therefore, there is a growing need to synthesize high-resolution and realistic CT scans for pediatric patients to improve deep learning organ segmentation.

Generative Adversarial Networks (GANs) \cite{C3} have gained a lot of attention due to their ability to generate realistic images from noise. The originally proposed GAN architecture consists of two convolutional neural networks competing with each other, namely a Generator network that generates synthetic images from noise and a Discriminator network that discriminates real samples from fake ones. Recently, new scalable GAN architectures successfully synthesizes high resolution images by progressively growing the resolution of both Generator and Discriminator layers \cite{C4}. However, these strategies are both computationally intensive and time-consuming. Moreover, medical images require highly precise and clinically meaningful structural information. Therefore, medical image synthesis is still at its novel stage due to major challenges presented in synthesizing viable and high resolution images. 

Little research has been conducted to conditionally synthesize realistic and high resolution medical images. Yang et. al. \cite{C5} uses a conditional GAN (cGAN) to perform cross-modality image-to-image translation and improve segmentation for Magnetic Resonance (MR) images. Costa et. al. \cite{C6} also reports a modified Pix2Pix GAN architecture which conditionally synthesizes retinal images given their respective vessel trees. As for CT image synthesis, Bowles et. al. \cite{C7} modified a Wasserstein GAN with training data reweighting to generate CT images modeling the progression of Alzheimer's disease. While all these methods are highly successful in synthesizing realistic medical images, the gradient becomes exponentially more unstable as GAN networks get deeper and deeper. This becomes a major limiting factor which prevents the successful synthesis of high resolution images. This is why our work provides an important foundation for the stable synthesis of high resolution CT images. 

The main focus of this study is to synthesize abdominal CT scans that contain the pancreas. CT images are synthesized conditionally according to the patients' ages. We choose Auxiliary Classifier Generative Adversarial Network (ACGAN) \cite{C8} as our base architecture due to its capability to incorporate label information into the GAN latent space. To increase our ACGAN's ability to learn the fine details in high resolution CT images without constructing a deep and unstable network, we propose the use of residual blocks in generator architecture. Residual blocks are proven to help stabilize deep generative adversarial networks, as previously shown in Bissoto et. al.'s work where they generated high resolution skin lesion images given instance and semantic maps \cite{C9}. Another technique we use to stabilize the gradient is the incorporation of pixelwise normalization layers \cite{C10}. By centering and dividing mini batches by their standard deviations, we can keep all the features in a similar range so the gradients do not go out of control. Finally, we specifically incorporate age information into our proposed network, since age can encode the organ shape variability by child growth. 
\begin{figure*}[t]
\begin{minipage}[b]{1.0\linewidth}
  \centering
  \centerline{\includegraphics[width=16cm]{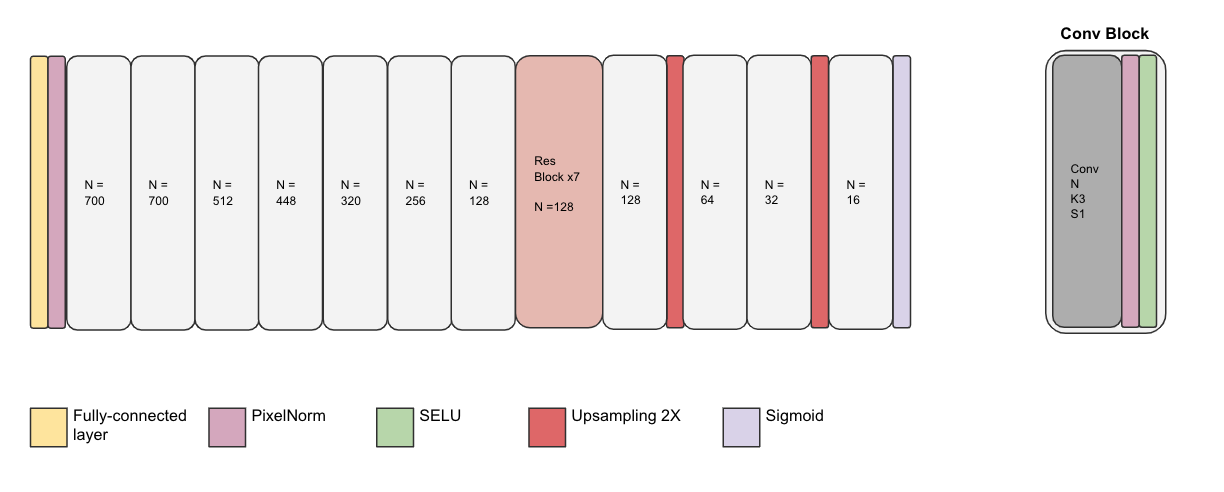}}
\end{minipage}
%\vspace{-30pt}
\caption{Overview of the Age-ACGAN generator architecture. Each convolutional block (denoted as Conv Block) consists of a convolutional layer, which is followed by a normalization (PixelNorm) and a scaled exponential linear unit (SELU) \cite{C11} activation layer. Each residual block is composed of 2 Conv Blocks with the same number of input and output feature maps. Despite having a different generator architecture, Age-ACGAN follows the same discriminator architecture as the original ACGAN.}
%\label{fig:res}
\end{figure*}

\section{Methods}
\label{sec:methodology}
\subsection{Generative Adversarial Network}
\label{ssec:gan}

The original GAN consists of two neural networks, namely a generator network $G$ and a discriminator network $D$ competing against each other. $G$ takes in a random noise vector $\mathbf{z}$, and transforms it into a generated image $G(\mathbf{z})$. The training objective of $D$ is to maximize $log(D(\mathbf{z})) + log(1-D(G(\mathbf{z}))$, the probability of assigning correct labels to both training images and images generated by the Generator . The $G$ network is trained to minimize $log(1-D(G(\mathbf{z}))$,  the log of the inverted probability of $D$ prediction of fake images. Since minimization of the inverted probability is not an easy task in practice, we seek to maximize the $D(G(\mathbf{z}))$ instead. In summary, the objective function of GAN can be formulated as a minimax loss:
\begin{multline}
\min_{\mathbf{G}}\max_{\mathbf{D}}V(D, G)=\mathbb{E}_{\mathbf{x}{\sim}p_{data}(\mathbf{x})}
[log D(\mathbf{x})] \\+ \mathbb{E}_{\mathbf{z}{\sim}p_{\mathbf{z}}(\mathbf{z})}
[log(1-D(G(\mathbf{z})))]
\end{multline}

This original formulation of GAN provides a powerful framework which aims to minimize the Jensen-Shannon divergence between the generated and real image data distributions. However, its architecture consists of simple fully connected layers and thereby rendering training unstable at times. 

Deep convolutional GANs (DCGANs)\cite{C12} are proposed to address instability issues in the original GAN architecture. It uses transposed convolutional layers instead of convolutional plus upsampling layers. DCGANs also introduce the use of batch normalization layers to improve convergence speed and training stability.

\subsection{Auxiliary Classifier GAN}
\label{ssec:acgan}

One major limitation of the original and deep convolution GANs is their inability to generate images conditionally. Conditional GANs are proposed to solve this issue by concatenating class labels $\mathbf{y}$ to random noise vector $\mathbf{z}$ as inputs to the generator network. Auxiliary classifier GAN (ACGAN) is a type of conditional GAN which produces images conditioned on their respective class labels. In addition to producing a probability distribution $P(S|X) = D(X)$ over possible images sources, the discriminator in ACGAN also produces a probability distribution $P(C|X) = D(X)$ over the class labels of the images. As a result, the objective function of the ACGAN can be defined as the log-likelihood of the correct source, $L_S$ and the log-likelihood of the correct class, $L_C$, where:
\begin{multline}
L_S = E[\log{P(S=real|X_{real})}] \\ +
E[\log{P(S=fake|X_{fake})}]
\end{multline}
\begin{multline}
L_C = E[\log{P(C=c|X_{real})}] \\ +
E[\log{P(C=c|X_{fake})}]
\end{multline}

\subsection{Age Conditioned Auxiliary Classifier GAN}
\label{ssec:objfunc}
We propose a variant of ACGAN architecture which we call an Age Auxiliary Classifier Generative Adversarial Network (Age-ACGAN) for the age-conditional synthesis of CT images. We modify the original objective function in ACGAN to compute the log-likelihoods of the correct source ($L_s$) and the correct age class ($L_a$) as following:
\vspace{-5pt}
\begin{multline}
L_s = E[\log{P(S_{CT}=real|X_{real})}] \\ +
E[\log{P(S_{CT}=fake|X_{fake})}]
\end{multline}
\begin{multline}
L_a = E[\log{P(C_{age}=age|X_{real})}] \\ +
E[\log{P(C_{age}=age|X_{fake})}]
\end{multline}

The training objective of our discriminator $D$ is to maximize $L_a + L_s$. This ensures the discriminator always maximizes the log likelihood it assigns to the correct source of CT image $CT_{source}$ and the correct age class $CT_{age}$.  

Even though the training objectives are kept similar, our proposed model has significant structural changes compared to the original ACGAN to accommodate the high resolution of medical images. (See Fig. 1.) The generator architecture of Age-ACGAN consists of 7 convolutional layers plus 9 residual blocks. We choose to use wider convolutional layers (up to 1024 channels) and smaller kernel sizes (3x3) than the original ACGAN architecture. Since the real images contain values between $0$ and $1$ and $tanh$ maps the output values to $[-1, 1]$, we change the last activation function in the generator network to $sigmoid$ instead. The discriminator of Age-ACGAN has one extra convolutional layer compared to the ACGAN's discriminator. Residual blocks in the generator are composed of two convolutional layers, SELU activation layers and PixelNorm normalization layers. In a traditional convolutional network, each layer is fed directly into the next layer. Degradation problems arise when networks get too deep and consist of too many layers. Residual blocks permit shortcuts for layers to feed directly to another layer deeper in the network. These skip connections allow us to construct a much deeper network without suffering from degradation issues to further capture the fine details in CT images with higher resolutions. \color{black} Additionally, the combination of SELU and PixelNorm layers allows the network to converge quickly and prevents gradient explosion. \color{black}

\section{Experimental Results}
\label{sec:experiments}

In order to test our proposed network's ability to conditionally generate CT images, we designed an experiment to generate abdominal CT images with segmentation masks. We synthesized the CT images along with their respective pancreas segmentation masks based on three age classes: infant class (ages 1 to 3), preschool class (ages 4 to 6) and adolescent class (ages 16 to 18).We used 20 slices from 5 patients for each of age class for training. We simply located the pancreas segmentation masks and cropped a bounding box around both the masks and the CT images. It is important to note that the images were not resized to lower resolutions to preserve high resolution details. 

We pre-processed CT images for intensity normalization and concatenated with their pancreas segmentation masks before being fed into our proposed network. We used cross entropy in our implementation to calculate the losses $L_s$ and $L_a$. During our synthesis of the segmentation masks, we chose an arbitrary cutoff point (e.g., $0.5$) to threshold synthesized masks values ranging from $[0, 1]$ to binary values $0$ and $1$. We were able to train the network with a fixed batch size of 16 for 3,000 epochs. 
% Specification of Cropping

\subsection{DCGAN vs. Age-ACGAN}
\label{ssec:The importance of conditional information}
We compare synthesis results of abdominal CT with pancreas segmentation masks from DCGAN and Age-ACGAN in Fig. 2. Our Age-ACGAN in Fig. 2 (b) improved the visual quality of the synthetic images compared with DCGAN in Fig. 2 (a). DCGAN shows the streaking artifact in synthesized CT image with irregular-shaped pancreas mask. On the other hand, our Age-ACGAN can synthesize realistic CT images and their respective masks in terms of CT noise texture and pancreas shape. This indicates that Age-ACGAN stabilizes the gradient when training the deeper network by incorporating age information.

\begin{figure}[t]
\begin{minipage}[b]{1.0\linewidth}
  \centering
  \centerline{\includegraphics[width=7cm, height=3.25cm]{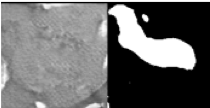}}
%  \vspace{2.0cm}
  \centerline{(a) DCGAN }\medskip
\end{minipage}
\begin{minipage}[b]{1.0\linewidth}
  \centering
  \centerline{\includegraphics[width=7cm, height=3.25cm]{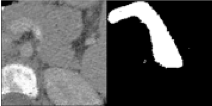}}
%  \vspace{2.0cm}
  \centerline{(b) Age-ACGAN}\medskip
\end{minipage}

\caption{Comparison of CT images with pancreas masks synthesized by (a) DCGAN and (b) Age-ACGAN.  Our Age-ACGAN is able to improve the visual quality of synthesized CT image and the corresponding pancreas masks compared to the DCGAN. Note the CT streak artifact and irregular pancreas shape in DCGAN synthesis while Age-ACGAN generating extremely fine detailed and high resolution images.}
\label{fig:res}
\end{figure}

For quantitative evaluation, we compare the convergence speed between the DCGAN and Age-ACGAN in Figure 3. Age-ACGAN converged after 500 iterations, while DCGAN failed to converge within the first 1000 iterations. We attribute the fast convergence speed of Age-ACGAN to the incorporation of age information, which enrich the information provided to both the generator and discriminator networks.

%%insert table here
\begin{figure}[t]
\begin{tikzpicture}
\begin{axis}[
    title={},
    xlabel={Iterations},
    ylabel={Generator Loss},
    xmin=0, xmax=1000,
    ymin=0, ymax=14,
    xtick={0,250, 500, 750, 1000},
    ytick={0,2,4,6,8,10, 12, 14},
    % legend pos=north west,
    ymajorgrids=true,
    grid style=dashed,
]
\addplot[
    color=red,
    ]
    coordinates {
(0,1.6823513507843018)(1,1.9865391254425049)(2,1.9929035902023315)(3,1.8897018432617188)(4,2.067009925842285)(5,2.3170220851898193)(6,2.9088916778564453)(7,2.484218120574951)(8,2.9128212928771973)(9,2.9155967235565186)(10,3.0627307891845703)(11,3.026923656463623)(12,3.437603235244751)(13,3.1337661743164062)(14,3.1896286010742188)(15,3.333378791809082)(16,3.262962818145752)(17,3.7484636306762695)(18,3.933802604675293)(19,3.8309054374694824)(20,3.5223283767700195)(21,3.6989355087280273)(22,4.129941463470459)(23,4.4787211418151855)(24,4.160447597503662)(25,3.748735189437866)(26,3.6238484382629395)(27,3.9793457984924316)(28,4.431031703948975)(29,4.545849800109863)(30,4.307029724121094)(31,4.417881965637207)(32,4.250452995300293)(33,4.319178104400635)(34,4.496239185333252)(35,4.521552085876465)(36,4.575582027435303)(37,4.683140754699707)(38,4.78190803527832)(39,4.927490234375)(40,4.900356292724609)(41,4.2057905197143555)(42,3.8941028118133545)(43,4.2363667488098145)(44,5.53603458404541)(45,5.535436153411865)(46,4.894530773162842)(47,5.028132915496826)(48,4.907801628112793)(49,4.8938984870910645)(50,4.677831172943115)(51,4.617181301116943)(52,4.606888771057129)(53,4.89581823348999)(54,4.856985092163086)(55,4.351498603820801)(56,4.713671684265137)(57,4.347270488739014)(58,4.964301586151123)(59,4.670476913452148)(60,4.685336589813232)(61,5.57529354095459)(62,5.155277729034424)(63,4.71691370010376)(64,4.045444488525391)(65,4.314329624176025)(66,4.7923903465271)(67,4.5587286949157715)(68,4.503490447998047)(69,4.913028717041016)(70,4.524794101715088)(71,4.5301971435546875)(72,4.4509124755859375)(73,4.767353534698486)(74,4.629031181335449)(75,4.447150707244873)(76,4.472805976867676)(77,4.561165809631348)(78,5.245772361755371)(79,4.758260250091553)(80,4.248061656951904)(81,4.713896751403809)(82,4.867220878601074)(83,4.600511074066162)(84,4.212874412536621)(85,4.69556188583374)(86,4.815474033355713)(87,4.609470367431641)(88,4.413854598999023)(89,4.022596836090088)(90,4.493197441101074)(91,4.4376420974731445)(92,4.473704814910889)(93,4.918456077575684)(94,5.225066661834717)(95,5.299511909484863)(96,5.028802394866943)(97,4.523432731628418)(98,4.411842346191406)(99,4.366074085235596)(100,4.455052375793457)(101,4.869707107543945)(102,5.218944549560547)(103,4.75013542175293)(104,4.340251445770264)(105,4.53220796585083)(106,4.923576354980469)(107,5.096827507019043)(108,5.061192989349365)(109,4.63145637512207)(110,4.435775279998779)(111,4.753304481506348)(112,4.844547271728516)(113,4.842705249786377)(114,4.757740020751953)(115,4.997176647186279)(116,5.044338703155518)(117,4.715452671051025)(118,5.129456996917725)(119,5.002961158752441)(120,5.1147003173828125)(121,5.5515851974487305)(122,4.8513007164001465)(123,4.902821063995361)(124,5.371093273162842)(125,4.97671365737915)(126,5.2569756507873535)(127,5.302760124206543)(128,4.955689430236816)(129,5.001436233520508)(130,5.253848075866699)(131,5.212122440338135)(132,5.422036647796631)(133,5.166320323944092)(134,5.356586456298828)(135,4.708361625671387)(136,5.063404560089111)(137,5.090286731719971)(138,5.388698577880859)(139,5.41438102722168)(140,5.485535144805908)(141,5.419602870941162)(142,4.815127372741699)(143,4.787441253662109)(144,4.841233253479004)(145,5.375161647796631)(146,5.832300186157227)(147,5.7983856201171875)(148,6.090034008026123)(149,5.6465535163879395)(150,5.035975933074951)(151,5.241034984588623)(152,5.470945358276367)(153,6.104589462280273)(154,6.387261867523193)(155,5.9472527503967285)(156,5.0027971267700195)(157,5.348333358764648)(158,5.225839614868164)(159,5.636499404907227)(160,6.2296576499938965)(161,5.505285739898682)(162,5.355398178100586)(163,6.343452453613281)(164,6.429296970367432)(165,5.96305513381958)(166,5.471534729003906)(167,5.022820472717285)(168,4.683189868927002)(169,5.2101593017578125)(170,5.60565710067749)(171,6.2281494140625)(172,6.245766639709473)(173,5.756035327911377)(174,6.029428958892822)(175,5.709573745727539)(176,5.528754234313965)(177,5.541281223297119)(178,5.1788010597229)(179,5.143061637878418)(180,5.860028266906738)(181,6.751133441925049)(182,6.433091640472412)(183,5.77461576461792)(184,5.396429538726807)(185,5.462464809417725)(186,5.594254970550537)(187,5.683250904083252)(188,5.953787803649902)(189,5.775257587432861)(190,5.723744869232178)(191,6.122648239135742)(192,6.147135257720947)(193,5.6991286277771)(194,5.892178058624268)(195,6.304356575012207)(196,6.937341690063477)(197,6.515049457550049)(198,5.566476345062256)(199,5.475386142730713)(200,5.983555316925049)(201,5.630504608154297)(202,6.1178083419799805)(203,7.207333087921143)(204,6.670176029205322)(205,7.017892837524414)(206,5.4232892990112305)(207,5.00657320022583)(208,5.0103631019592285)(209,5.06707763671875)(210,5.518632411956787)(211,5.846176624298096)(212,5.984523296356201)(213,6.254175186157227)(214,7.1879143714904785)(215,7.602747917175293)(216,6.925369739532471)(217,5.689704418182373)(218,5.057310104370117)(219,4.839285850524902)(220,6.518612861633301)(221,6.3991498947143555)(222,6.674422264099121)(223,5.8908843994140625)(224,5.965758800506592)(225,5.970570087432861)(226,5.903191089630127)(227,5.448431968688965)(228,5.592826843261719)(229,5.488805294036865)(230,6.019843101501465)(231,6.064457893371582)(232,6.0816755294799805)(233,6.292938232421875)(234,6.305704116821289)(235,6.403512954711914)(236,6.100235939025879)(237,6.454353332519531)(238,6.527892589569092)(239,6.529790878295898)(240,6.251436710357666)(241,6.419644355773926)(242,6.293488502502441)(243,6.631872653961182)(244,6.064460754394531)(245,6.037298202514648)(246,6.076478958129883)(247,6.037655830383301)(248,6.235609531402588)(249,6.511348724365234)(250,6.168407917022705)(251,6.174406051635742)(252,6.211702346801758)(253,6.112740516662598)(254,6.22899055480957)(255,5.941471099853516)(256,5.981705188751221)(257,6.317017555236816)(258,6.392332077026367)(259,6.336384296417236)(260,6.305468559265137)(261,6.665184497833252)(262,6.376209259033203)(263,6.668939113616943)(264,6.234314918518066)(265,6.092589378356934)(266,6.383458137512207)(267,6.9935736656188965)(268,7.094756126403809)(269,7.67498254776001)(270,7.635953426361084)(271,7.190516948699951)(272,6.834531784057617)(273,6.588023662567139)(274,6.47671365737915)(275,6.145634174346924)(276,5.545468807220459)(277,5.984827518463135)(278,6.168938636779785)(279,6.356968879699707)(280,6.43338680267334)(281,6.465983867645264)(282,6.6289448738098145)(283,6.931954383850098)(284,6.885656833648682)(285,7.2846503257751465)(286,7.423924446105957)(287,6.782405853271484)(288,6.658337593078613)(289,6.2757086753845215)(290,6.026968479156494)(291,6.294917583465576)(292,6.243645191192627)(293,6.498898506164551)(294,6.599797248840332)(295,6.872394561767578)(296,6.782575607299805)(297,6.532496929168701)(298,6.752806663513184)(299,6.795048713684082)(300,6.684562683105469)(301,6.514720439910889)(302,6.418527126312256)(303,6.734790802001953)(304,6.4231486320495605)(305,6.474246501922607)(306,6.742778778076172)(307,7.6356682777404785)(308,7.095033168792725)(309,6.426877975463867)(310,6.84408712387085)(311,7.024354934692383)(312,6.602602481842041)(313,6.525029182434082)(314,6.399835586547852)(315,6.074186325073242)(316,6.567169189453125)(317,6.530081748962402)(318,6.390533447265625)(319,6.518243789672852)(320,6.742305755615234)(321,6.939640522003174)(322,6.915187358856201)(323,7.318093776702881)(324,7.390857696533203)(325,7.380032539367676)(326,6.627633094787598)(327,6.516922950744629)(328,6.604881763458252)(329,7.016483783721924)(330,7.27241849899292)(331,7.308462619781494)(332,6.733808517456055)(333,6.381969451904297)(334,6.136053085327148)(335,6.202216625213623)(336,6.32650089263916)(337,6.762516498565674)(338,6.263509273529053)(339,6.464996814727783)(340,7.3882317543029785)(341,7.016873359680176)(342,6.877635478973389)(343,6.7674641609191895)(344,6.7505645751953125)(345,7.19771671295166)(346,6.926052093505859)(347,9.335251808166504)(348,5.807003021240234)(349,4.573188781738281)(350,6.2838592529296875)(351,7.035884380340576)(352,6.985101699829102)(353,7.067252159118652)(354,6.833799839019775)(355,6.743657112121582)(356,6.605765342712402)(357,6.606468200683594)(358,6.71163272857666)(359,6.530762672424316)(360,6.7688212394714355)(361,6.805771827697754)(362,6.892033576965332)(363,6.629823684692383)(364,6.789697647094727)(365,6.939360618591309)(366,6.8542985916137695)(367,7.01193904876709)(368,6.801711559295654)(369,6.8683953285217285)(370,7.449367046356201)(371,8.590866088867188)(372,7.920069694519043)(373,7.969291687011719)(374,8.22169017791748)(375,7.223630428314209)(376,6.7501654624938965)(377,6.882524013519287)(378,6.858497142791748)(379,6.602550506591797)(380,6.730912208557129)(381,6.7765278816223145)(382,7.00198221206665)(383,7.009028434753418)(384,6.570198059082031)(385,7.242953300476074)(386,7.693295955657959)(387,6.9661455154418945)(388,6.951977729797363)(389,6.806969165802002)(390,7.159990310668945)(391,7.433497905731201)(392,7.186361312866211)(393,6.704607963562012)(394,7.056805610656738)(395,7.059452533721924)(396,6.985964298248291)(397,6.747795581817627)(398,6.9239277839660645)(399,7.166468143463135)(400,7.2648606300354)(401,7.053062438964844)(402,7.137044906616211)(403,7.465475559234619)(404,7.56173849105835)(405,7.732648849487305)(406,8.056958198547363)(407,8.06738567352295)(408,7.317734718322754)(409,7.032871723175049)(410,7.043030738830566)(411,7.046096324920654)(412,7.157548904418945)(413,7.0548810958862305)(414,6.870240211486816)(415,7.44355583190918)(416,7.240472316741943)(417,7.460687160491943)(418,8.070740699768066)(419,8.246387481689453)(420,7.512868404388428)(421,6.518901348114014)(422,5.99179220199585)(423,6.446189880371094)(424,6.540403366088867)(425,6.790586471557617)(426,6.990013599395752)(427,7.7016215324401855)(428,7.39599084854126)(429,6.886806488037109)(430,7.833773136138916)(431,8.319181442260742)(432,6.787502288818359)(433,5.5526041984558105)(434,6.361305236816406)(435,7.375429153442383)(436,7.501408100128174)(437,7.229673862457275)(438,6.496282577514648)(439,6.6190290451049805)(440,7.188932418823242)(441,7.132307052612305)(442,6.27810001373291)(443,7.172698020935059)(444,6.797098636627197)(445,6.705583572387695)(446,6.826828479766846)(447,6.709347248077393)(448,8.014924049377441)(449,6.838442325592041)(450,6.1015238761901855)(451,7.334073543548584)(452,7.0086870193481445)(453,6.841994285583496)(454,6.688863277435303)(455,7.26141357421875)(456,6.463035583496094)(457,6.306936740875244)(458,9.381829261779785)(459,12.6299467086792)(460,7.051522254943848)(461,5.066910266876221)(462,8.110658645629883)(463,9.587024688720703)(464,9.963130950927734)(465,9.497791290283203)(466,9.003743171691895)(467,8.832534790039062)(468,8.101116180419922)(469,7.2877631187438965)(470,7.383255958557129)(471,7.198374271392822)(472,7.070910453796387)(473,7.590755939483643)(474,7.375004768371582)(475,7.417749881744385)(476,7.0348591804504395)(477,6.847724437713623)(478,6.929483413696289)(479,7.303618431091309)(480,7.049123764038086)(481,7.319945812225342)(482,7.765066623687744)(483,8.568095207214355)(484,7.260345935821533)(485,7.607137203216553)(486,7.904278755187988)(487,7.510912895202637)(488,7.4876275062561035)(489,7.4825849533081055)(490,7.157965183258057)(491,7.313360214233398)(492,7.706917762756348)(493,8.081191062927246)(494,7.897243976593018)(495,7.698216915130615)(496,7.1981730461120605)(497,6.929163932800293)(498,7.284809112548828)(499,7.438847541809082)(500,7.502350807189941)(501,7.77610969543457)(502,8.0389404296875)(503,8.432640075683594)(504,8.17255973815918)(505,7.657589912414551)(506,7.700891971588135)(507,7.905696868896484)(508,7.987810134887695)(509,7.624492168426514)(510,7.4161224365234375)(511,7.553417205810547)(512,7.441948413848877)(513,7.139169692993164)(514,7.723719596862793)(515,7.529937744140625)(516,7.610453128814697)(517,8.184054374694824)(518,8.270939826965332)(519,7.944090843200684)(520,7.255258560180664)(521,7.5779829025268555)(522,7.024075508117676)(523,7.881961822509766)(524,7.701541423797607)(525,7.240411281585693)(526,6.977675437927246)(527,7.0056891441345215)(528,8.452902793884277)(529,8.927719116210938)(530,8.278944969177246)(531,7.540053367614746)(532,7.415029525756836)(533,7.820857048034668)(534,6.910763740539551)(535,7.385637283325195)(536,7.236978530883789)(537,7.2531280517578125)(538,6.821702480316162)(539,6.091484069824219)(540,9.040754318237305)(541,7.724539756774902)(542,10.691020011901855)(543,8.759754180908203)(544,6.375187873840332)(545,5.888288497924805)(546,5.745745658874512)(547,7.268958568572998)(548,8.005391120910645)(549,7.458652019500732)(550,7.775990962982178)(551,13.348321914672852)(552,10.783097267150879)(553,9.320180892944336)(554,8.628389358520508)(555,8.330714225769043)(556,8.025636672973633)(557,7.836586952209473)(558,7.388090133666992)(559,7.4335856437683105)(560,7.131515026092529)(561,7.326323509216309)(562,8.961122512817383)(563,9.251747131347656)(564,8.5614013671875)(565,7.826288223266602)(566,7.554582118988037)(567,7.519731044769287)(568,7.479833602905273)(569,7.476188659667969)(570,7.553414821624756)(571,7.377892017364502)(572,7.833597660064697)(573,7.323085784912109)(574,7.772342681884766)(575,7.828920364379883)(576,7.312158107757568)(577,7.516204833984375)(578,7.456846714019775)(579,7.529049396514893)(580,7.2440104484558105)(581,7.052128791809082)(582,7.260855197906494)(583,7.224187850952148)(584,6.81728458404541)(585,7.129670143127441)(586,7.333302021026611)(587,7.12255859375)(588,7.045855522155762)(589,7.354637622833252)(590,7.544404983520508)(591,7.929762840270996)(592,8.447466850280762)(593,8.14990520477295)(594,6.616359710693359)(595,6.367353439331055)(596,6.852136611938477)(597,7.2051801681518555)(598,6.635655403137207)(599,7.836789608001709)(600,8.390695571899414)(601,8.0180025100708)(602,8.197509765625)(603,7.843514919281006)(604,7.865742206573486)(605,7.980733871459961)(606,7.525362491607666)(607,6.826473712921143)(608,7.162656784057617)(609,7.291464328765869)(610,7.505923271179199)(611,7.484277725219727)(612,7.67067813873291)(613,7.751493453979492)(614,7.28708553314209)(615,7.515913963317871)(616,7.176163673400879)(617,7.567564964294434)(618,7.6285223960876465)(619,7.810387134552002)(620,7.816920757293701)(621,7.511563777923584)(622,7.861502170562744)(623,7.493575096130371)(624,7.225904941558838)(625,7.065130710601807)(626,7.199193477630615)(627,7.955533027648926)(628,7.939672946929932)(629,8.34740924835205)(630,7.634091854095459)(631,8.042728424072266)(632,7.28043270111084)(633,6.857834815979004)(634,7.072286605834961)(635,8.326708793640137)(636,8.559000015258789)(637,8.7008638381958)(638,8.57332706451416)(639,8.057042121887207)(640,8.224079132080078)(641,7.773680210113525)(642,7.096436023712158)(643,7.250565052032471)(644,6.967960834503174)(645,7.346076011657715)(646,7.26613712310791)(647,7.17045783996582)(648,7.4015021324157715)(649,7.507466793060303)(650,7.656944274902344)(651,7.859245777130127)(652,7.576339244842529)(653,7.566708087921143)(654,7.43605375289917)(655,7.579992771148682)(656,7.822803497314453)(657,7.767906188964844)(658,7.40266227722168)(659,6.917633056640625)(660,7.287808418273926)(661,6.911071300506592)(662,7.5246901512146)(663,7.723260402679443)(664,8.172475814819336)(665,7.438190460205078)(666,7.573822498321533)(667,7.381722450256348)(668,7.643679618835449)(669,8.171058654785156)(670,7.7446136474609375)(671,7.7130022048950195)(672,8.094167709350586)(673,8.02491283416748)(674,7.857842445373535)(675,8.047852516174316)(676,8.225050926208496)(677,8.36188793182373)(678,8.1782865524292)(679,8.228850364685059)(680,8.038293838500977)(681,7.96309232711792)(682,8.260797500610352)(683,8.191142082214355)(684,8.114045143127441)(685,7.964459419250488)(686,8.039859771728516)(687,8.174361228942871)(688,7.977563858032227)(689,7.990437984466553)(690,8.103734970092773)(691,7.95295524597168)(692,8.128386497497559)(693,7.898695945739746)(694,8.31579875946045)(695,8.375864028930664)(696,8.232271194458008)(697,8.165867805480957)(698,8.1544771194458)(699,8.216180801391602)(700,8.117527961730957)(701,8.106459617614746)(702,8.276131629943848)(703,8.312828063964844)(704,7.611950397491455)(705,6.884987831115723)(706,7.054635047912598)(707,6.91289758682251)(708,7.320606231689453)(709,7.519097328186035)(710,7.322463035583496)(711,7.3497314453125)(712,7.438080310821533)(713,7.653789520263672)(714,7.535429000854492)(715,7.7518534660339355)(716,7.6971821784973145)(717,7.811429023742676)(718,7.826849937438965)(719,7.681040287017822)(720,8.115501403808594)(721,7.90020751953125)(722,8.106954574584961)(723,8.16096019744873)(724,8.14830207824707)(725,7.8633904457092285)(726,8.14072036743164)(727,8.052637100219727)(728,7.965756893157959)(729,8.16293716430664)(730,8.337187767028809)(731,8.362991333007812)(732,8.16781997680664)(733,8.886672973632812)(734,8.438529014587402)(735,8.13441276550293)(736,8.413627624511719)(737,8.730870246887207)(738,8.255908966064453)(739,8.330681800842285)(740,8.957747459411621)(741,8.540513038635254)(742,8.648554801940918)(743,8.3854341506958)(744,8.97976303100586)(745,8.731724739074707)(746,8.407981872558594)(747,8.539255142211914)(748,8.610820770263672)(749,8.111088752746582)(750,7.899579048156738)(751,8.384350776672363)(752,7.9710798263549805)(753,7.887025356292725)(754,7.814764499664307)(755,7.783412933349609)(756,7.899383544921875)(757,7.94980525970459)(758,8.086259841918945)(759,8.375744819641113)(760,8.450993537902832)(761,8.013604164123535)(762,8.349817276000977)(763,8.193597793579102)(764,8.277107238769531)(765,8.230545997619629)(766,7.9486517906188965)(767,8.03754997253418)(768,8.42086124420166)(769,8.296649932861328)(770,8.333232879638672)(771,8.188467979431152)(772,8.142997741699219)(773,8.24878978729248)(774,8.3143310546875)(775,8.482532501220703)(776,8.509794235229492)(777,8.715041160583496)(778,8.169206619262695)(779,8.404671669006348)(780,8.302014350891113)(781,8.697159767150879)(782,8.577723503112793)(783,8.567540168762207)(784,8.871077537536621)(785,8.495213508605957)(786,8.503389358520508)(787,8.173188209533691)(788,8.130694389343262)(789,8.284671783447266)(790,8.204866409301758)(791,8.51866340637207)(792,8.77140998840332)(793,8.62502384185791)(794,8.579469680786133)(795,8.449727058410645)(796,8.753877639770508)(797,8.47075080871582)(798,8.208227157592773)(799,8.45870590209961)(800,8.228070259094238)(801,8.340904235839844)(802,8.544759750366211)(803,8.587130546569824)(804,8.354958534240723)(805,8.530258178710938)(806,8.369166374206543)(807,8.867405891418457)(808,8.295984268188477)(809,8.46545696258545)(810,8.454483985900879)(811,8.29946231842041)(812,8.271230697631836)(813,8.570910453796387)(814,8.827143669128418)(815,8.723526954650879)(816,8.678607940673828)(817,8.704947471618652)(818,8.560808181762695)(819,8.537213325500488)(820,8.704317092895508)(821,8.640188217163086)(822,8.467146873474121)(823,8.673491477966309)(824,8.400894165039062)(825,8.658805847167969)(826,8.797743797302246)(827,8.505236625671387)(828,8.550601959228516)(829,8.67361831665039)(830,8.879444122314453)(831,8.677412986755371)(832,8.6878662109375)(833,8.979605674743652)(834,9.434678077697754)(835,8.641857147216797)(836,8.415804862976074)(837,8.357514381408691)(838,8.55844497680664)(839,8.88322639465332)(840,8.54293441772461)(841,8.912853240966797)(842,8.778661727905273)(843,9.351583480834961)(844,8.59130859375)(845,8.754044532775879)(846,8.676011085510254)(847,8.531495094299316)(848,8.662781715393066)(849,8.590203285217285)(850,8.628142356872559)(851,8.633421897888184)(852,8.442012786865234)(853,8.520265579223633)(854,8.384246826171875)(855,8.636909484863281)(856,8.797359466552734)(857,8.606592178344727)(858,9.008641242980957)(859,8.857684135437012)(860,8.947704315185547)(861,8.846536636352539)(862,8.821993827819824)(863,8.724468231201172)(864,8.735674858093262)(865,8.745590209960938)(866,8.944794654846191)(867,8.87187385559082)(868,8.784492492675781)(869,8.819523811340332)(870,8.738727569580078)(871,9.825628280639648)(872,9.753931045532227)(873,9.780769348144531)(874,9.926105499267578)(875,9.966290473937988)(876,9.667339324951172)(877,9.552850723266602)(878,9.462179183959961)(879,9.198031425476074)(880,9.385138511657715)(881,9.238170623779297)(882,9.285335540771484)(883,10.05887508392334)(884,9.632269859313965)(885,8.951886177062988)(886,9.393646240234375)(887,8.959229469299316)(888,8.494058609008789)(889,8.45821762084961)(890,8.259160041809082)(891,8.411200523376465)(892,8.200688362121582)(893,7.96903133392334)(894,8.29415225982666)(895,8.787647247314453)(896,8.50446891784668)(897,8.971060752868652)(898,8.60661506652832)(899,8.749046325683594)(900,8.49746036529541)(901,8.42442512512207)(902,8.858084678649902)(903,9.01488208770752)(904,8.86194133758545)(905,8.62134075164795)(906,8.881715774536133)(907,8.907377243041992)(908,8.477112770080566)(909,8.735722541809082)(910,8.994763374328613)(911,8.728508949279785)(912,8.876057624816895)(913,8.79539966583252)(914,8.887261390686035)(915,8.59247875213623)(916,8.957176208496094)(917,8.698537826538086)(918,9.028059959411621)(919,9.014202117919922)(920,9.107244491577148)(921,9.020623207092285)(922,9.033323287963867)(923,8.647692680358887)(924,8.729635238647461)(925,9.15664005279541)(926,9.36940860748291)(927,9.211103439331055)(928,8.752826690673828)(929,9.587291717529297)(930,9.430431365966797)(931,9.089771270751953)(932,9.097403526306152)(933,9.118681907653809)(934,8.839875221252441)(935,9.191699028015137)(936,8.715189933776855)(937,9.072543144226074)(938,9.204131126403809)(939,8.668959617614746)(940,8.937901496887207)(941,8.91721248626709)(942,8.860522270202637)(943,8.82008171081543)(944,9.18683910369873)(945,8.797781944274902)(946,8.970736503601074)(947,8.94521427154541)(948,8.899802207946777)(949,8.660124778747559)(950,8.844948768615723)(951,8.836830139160156)(952,8.826754570007324)(953,9.166160583496094)(954,9.225488662719727)(955,9.336515426635742)(956,8.441571235656738)(957,8.248127937316895)(958,8.438878059387207)(959,8.279258728027344)(960,8.398086547851562)(961,8.293495178222656)(962,8.636741638183594)(963,8.8068265914917)(964,8.613158226013184)(965,8.937054634094238)(966,8.6823091506958)(967,8.955936431884766)(968,9.344215393066406)(969,9.329530715942383)(970,9.021598815917969)(971,9.269922256469727)(972,8.747661590576172)(973,9.124828338623047)(974,9.189501762390137)(975,9.0663423538208)(976,9.094287872314453)(977,9.520594596862793)(978,8.969101905822754)(979,9.434350967407227)(980,9.893620491027832)(981,9.445294380187988)(982,9.164249420166016)(983,9.176535606384277)(984,8.949464797973633)(985,8.78048038482666)(986,9.014808654785156)(987,8.926148414611816)(988,8.817323684692383)(989,8.693029403686523)(990,8.970962524414062)(991,8.72116756439209)(992,9.176077842712402)(993,9.012076377868652)(994,9.008369445800781)(995,9.248235702514648)(996,9.182519912719727)(997,8.461642265319824)(998,9.069439888000488)(999,9.011402130126953)(1000,8.839278221130371)
    };
 
\addplot[
    color=blue,
    ]
    coordinates {
    (0,1.5052244663238525)(1,1.5332894325256348)(2,1.5254333019256592)(3,1.4732258319854736)(4,1.3665294647216797)(5,1.4441428184509277)(6,1.370059609413147)(7,1.2373197078704834)(8,1.4080801010131836)(9,1.432441234588623)(10,1.309079647064209)(11,1.1666967868804932)(12,1.5508875846862793)(13,1.4277520179748535)(14,1.3499338626861572)(15,1.3972021341323853)(16,1.243638038635254)(17,1.5362012386322021)(18,1.3774099349975586)(19,1.427180528640747)(20,1.1971800327301025)(21,1.2396060228347778)(22,1.2139931917190552)(23,1.1036288738250732)(24,1.3030340671539307)(25,1.100790023803711)(26,1.154548168182373)(27,0.9988072514533997)(28,1.444638967514038)(29,1.4414887428283691)(30,1.6083722114562988)(31,1.4790799617767334)(32,1.2678430080413818)(33,1.068965196609497)(34,1.073150634765625)(35,1.2250944375991821)(36,1.0936050415039062)(37,1.352504014968872)(38,1.1938745975494385)(39,1.279840350151062)(40,1.2672353982925415)(41,1.265256404876709)(42,1.1528339385986328)(43,1.4239462614059448)(44,1.0195024013519287)(45,1.1878433227539062)(46,1.480949878692627)(47,1.5355277061462402)(48,1.4963382482528687)(49,1.4023866653442383)(50,1.4688423871994019)(51,1.3186463117599487)(52,1.2485709190368652)(53,1.2508583068847656)(54,1.3373247385025024)(55,1.861154556274414)(56,1.513358235359192)(57,1.7143205404281616)(58,1.5216450691223145)(59,1.3984793424606323)(60,2.083164691925049)(61,1.574896216392517)(62,2.013777256011963)(63,1.6946589946746826)(64,1.7454363107681274)(65,1.727763295173645)(66,1.3778612613677979)(67,1.868956446647644)(68,1.556287407875061)(69,2.5251379013061523)(70,1.8871774673461914)(71,2.892880916595459)(72,2.1986629962921143)(73,1.8957054615020752)(74,2.526865243911743)(75,1.9270501136779785)(76,1.8246537446975708)(77,1.5405254364013672)(78,2.0157995223999023)(79,1.9082249402999878)(80,2.3177809715270996)(81,0.7542864680290222)(82,1.3054988384246826)(83,2.8723058700561523)(84,1.351354718208313)(85,1.6497405767440796)(86,2.146731376647949)(87,1.9079618453979492)(88,1.6852492094039917)(89,1.9461867809295654)(90,2.0446064472198486)(91,1.416986346244812)(92,1.8712745904922485)(93,1.985297441482544)(94,1.3705002069473267)(95,2.087829113006592)(96,1.9238749742507935)(97,2.029700994491577)(98,1.1395864486694336)(99,2.9452648162841797)(100,0.25262385606765747)(101,0.5353552103042603)(102,1.0754140615463257)(103,1.9066938161849976)(104,2.109825611114502)(105,1.578782320022583)(106,1.1692194938659668)(107,1.1631656885147095)(108,1.9159034490585327)(109,2.956922769546509)(110,1.6211285591125488)(111,2.141960859298706)(112,2.3072988986968994)(113,1.6472238302230835)(114,1.41085946559906)(115,1.905844807624817)(116,1.5816799402236938)(117,1.7012354135513306)(118,2.33203387260437)(119,2.17482852935791)(120,1.3318220376968384)(121,2.124903440475464)(122,2.349339246749878)(123,1.7044562101364136)(124,2.2822952270507812)(125,2.1561124324798584)(126,1.703897476196289)(127,1.149492859840393)(128,1.6602351665496826)(129,1.5654760599136353)(130,1.1340943574905396)(131,2.104142665863037)(132,2.096024990081787)(133,1.8383352756500244)(134,2.8851025104522705)(135,2.7647385597229004)(136,2.289020299911499)(137,3.4492690563201904)(138,2.3365561962127686)(139,2.144265651702881)(140,2.943474292755127)(141,1.4064900875091553)(142,2.149705410003662)(143,2.3097665309906006)(144,2.3778862953186035)(145,2.1561150550842285)(146,3.7399990558624268)(147,3.6201884746551514)(148,3.754426956176758)(149,2.9168455600738525)(150,1.9656661748886108)(151,2.616849422454834)(152,2.7722983360290527)(153,2.810325860977173)(154,4.3568925857543945)(155,3.618786334991455)(156,5.536022186279297)(157,2.9727165699005127)(158,3.027858257293701)(159,3.1532816886901855)(160,3.0951690673828125)(161,2.348714828491211)(162,1.4362996816635132)(163,3.6069741249084473)(164,0.301172137260437)(165,1.4606434106826782)(166,3.5632338523864746)(167,3.4190785884857178)(168,2.000192403793335)(169,1.9237834215164185)(170,3.2821695804595947)(171,3.0872604846954346)(172,2.461158275604248)(173,2.873410940170288)(174,2.4898464679718018)(175,2.90281343460083)(176,2.246964931488037)(177,2.9943134784698486)(178,3.3251352310180664)(179,2.814906597137451)(180,2.503748655319214)(181,2.301800489425659)(182,2.018280506134033)(183,2.328291893005371)(184,1.6267127990722656)(185,1.7724642753601074)(186,3.3529250621795654)(187,1.6147516965866089)(188,2.6633269786834717)(189,0.4153684675693512)(190,2.268829107284546)(191,2.677208662033081)(192,0.9422823786735535)(193,2.2543139457702637)(194,2.2392144203186035)(195,1.192775845527649)(196,1.884955644607544)(197,2.6140294075012207)(198,1.092640995979309)(199,2.5174782276153564)(200,1.6561132669448853)(201,1.489423394203186)(202,1.695753812789917)(203,3.936270236968994)(204,0.2569168508052826)(205,3.6192574501037598)(206,3.3379595279693604)(207,2.5606987476348877)(208,1.7268401384353638)(209,3.240342140197754)(210,1.139623761177063)(211,1.7449660301208496)(212,2.2691256999969482)(213,3.294973850250244)(214,2.576096534729004)(215,2.996354341506958)(216,1.1746177673339844)(217,6.0779619216918945)(218,0.23710942268371582)(219,0.8269391059875488)(220,1.5618088245391846)(221,2.88284969329834)(222,1.8167386054992676)(223,1.149369239807129)(224,1.1847565174102783)(225,1.9963277578353882)(226,1.0124648809432983)(227,1.1092238426208496)(228,1.5034761428833008)(229,1.32306969165802)(230,1.6594171524047852)(231,1.5393670797348022)(232,1.4038091897964478)(233,1.4029641151428223)(234,2.6287431716918945)(235,0.5259582996368408)(236,3.1082215309143066)(237,1.2780773639678955)(238,1.5217045545578003)(239,1.763992428779602)(240,1.7540156841278076)(241,1.370940923690796)(242,2.3261783123016357)(243,3.106790781021118)(244,1.1091848611831665)(245,2.43129301071167)(246,2.1488757133483887)(247,1.614151954650879)(248,2.1477863788604736)(249,2.104966163635254)(250,0.8806737661361694)(251,4.526598930358887)(252,0.17071160674095154)(253,1.497573971748352)(254,2.3814821243286133)(255,0.9605132341384888)(256,1.3308320045471191)(257,2.367666006088257)(258,0.9527625441551208)(259,1.9100342988967896)(260,1.3677088022232056)(261,2.8814425468444824)(262,0.9216657876968384)(263,2.116914749145508)(264,3.4265339374542236)(265,0.6166890859603882)(266,1.8193809986114502)(267,2.4793434143066406)(268,3.1695334911346436)(269,2.554067850112915)(270,1.656071662902832)(271,2.2537126541137695)(272,0.8790517449378967)(273,3.279496669769287)(274,2.779400587081909)(275,0.3756837248802185)(276,5.791261672973633)(277,0.7444299459457397)(278,1.0866410732269287)(279,2.621673345565796)(280,3.4700536727905273)(281,1.7398388385772705)(282,2.7556655406951904)(283,0.917036235332489)(284,2.1028592586517334)(285,2.3634324073791504)(286,1.7666206359863281)(287,1.2192797660827637)(288,2.1505143642425537)(289,0.6857824921607971)(290,3.0810627937316895)(291,0.2770497500896454)(292,3.689286231994629)(293,0.9673124551773071)(294,1.2344869375228882)(295,2.3614399433135986)(296,1.814179539680481)(297,1.0031046867370605)(298,2.3643267154693604)(299,3.027336835861206)(300,0.82368004322052)(301,1.3005503416061401)(302,2.465083360671997)(303,1.0075228214263916)(304,1.915254831314087)(305,2.012079954147339)(306,2.1338798999786377)(307,2.567708969116211)(308,3.103513240814209)(309,1.3560643196105957)(310,3.6119608879089355)(311,0.6593765020370483)(312,1.756720781326294)(313,3.288516044616699)(314,0.6893073916435242)(315,1.2144795656204224)(316,2.8301055431365967)(317,1.2609630823135376)(318,1.311642050743103)(319,1.9843097925186157)(320,1.2072205543518066)(321,1.4203771352767944)(322,2.2941954135894775)(323,1.048447847366333)(324,3.097191572189331)(325,0.5374565720558167)(326,3.2081243991851807)(327,1.5788099765777588)(328,1.5957646369934082)(329,2.8917646408081055)(330,3.1646523475646973)(331,1.7822915315628052)(332,2.060215473175049)(333,1.2546918392181396)(334,2.756981372833252)(335,1.3160655498504639)(336,3.185257911682129)(337,2.5537478923797607)(338,1.9188076257705688)(339,2.1223065853118896)(340,3.098458766937256)(341,2.0365188121795654)(342,2.153583288192749)(343,0.5605869293212891)(344,2.56587290763855)(345,1.9946705102920532)(346,1.8404814004898071)(347,1.5924142599105835)(348,3.0664587020874023)(349,3.345115900039673)(350,0.9350802898406982)(351,3.5872182846069336)(352,3.448342800140381)(353,3.340576171875)(354,2.034106492996216)(355,2.9302642345428467)(356,4.484533309936523)(357,3.6330175399780273)(358,2.344419479370117)(359,2.0696587562561035)(360,3.3153138160705566)(361,5.116036415100098)(362,5.127223014831543)(363,3.29367733001709)(364,3.8402562141418457)(365,4.462909698486328)(366,5.559174060821533)(367,6.540678977966309)(368,7.855740070343018)(369,7.245768070220947)(370,6.030850887298584)(371,5.601452350616455)(372,4.369898796081543)(373,5.752291202545166)(374,5.572458744049072)(375,4.488844394683838)(376,5.474884510040283)(377,5.182577133178711)(378,4.623633861541748)(379,5.128292083740234)(380,5.209253311157227)(381,5.8291168212890625)(382,5.884662628173828)(383,4.601225852966309)(384,5.014944553375244)(385,5.332400321960449)(386,6.048422336578369)(387,6.524191856384277)(388,6.227772235870361)(389,6.83524751663208)(390,8.071952819824219)(391,7.329940319061279)(392,5.437614440917969)(393,3.9252543449401855)(394,3.891400098800659)(395,4.0160088539123535)(396,5.738065242767334)(397,5.280491352081299)(398,7.71274471282959)(399,8.506659507751465)(400,7.814844608306885)(401,6.988219261169434)(402,6.82751989364624)(403,5.89286470413208)(404,5.091230869293213)(405,5.047472953796387)(406,4.542843341827393)(407,4.307394504547119)(408,4.386248588562012)(409,4.31697940826416)(410,4.432284832000732)(411,6.090904712677002)(412,8.27303409576416)(413,8.37561321258545)(414,7.7437052726745605)(415,7.098226547241211)(416,6.873424530029297)(417,5.488458156585693)(418,5.0448760986328125)(419,4.594517707824707)(420,5.176353454589844)(421,4.3325605392456055)(422,4.465371131896973)(423,7.004810333251953)(424,7.701291561126709)(425,7.608946800231934)(426,5.671331405639648)(427,5.128515720367432)(428,4.877256870269775)(429,4.460127830505371)(430,4.251523971557617)(431,4.2332258224487305)(432,4.544020652770996)(433,5.45359992980957)(434,7.0964460372924805)(435,6.45845365524292)(436,5.698572158813477)(437,4.645129203796387)(438,4.677003860473633)(439,3.8420207500457764)(440,4.536362648010254)(441,4.268640995025635)(442,7.187960624694824)(443,6.753218173980713)(444,6.420432090759277)(445,5.90970516204834)(446,5.277660846710205)(447,4.9787797927856445)(448,4.7756171226501465)(449,4.763383388519287)(450,3.868718147277832)(451,4.58178186416626)(452,7.436257839202881)(453,6.522594928741455)(454,6.935689449310303)(455,5.75890588760376)(456,5.384604454040527)(457,4.586244583129883)(458,4.510964870452881)(459,4.136329174041748)(460,4.208853721618652)(461,4.489538669586182)(462,4.835544109344482)(463,4.095975399017334)(464,5.250519275665283)(465,4.472569942474365)(466,5.898598670959473)(467,3.8304898738861084)(468,3.5469515323638916)(469,3.9619624614715576)(470,4.051212787628174)(471,5.51488733291626)(472,2.8927323818206787)(473,4.343930244445801)(474,2.8778228759765625)(475,2.748775005340576)(476,3.0701005458831787)(477,3.241950035095215)(478,2.8711299896240234)(479,2.798874855041504)(480,2.8260104656219482)(481,3.2131130695343018)(482,2.9605729579925537)(483,2.977583408355713)(484,2.8562121391296387)(485,2.8146700859069824)(486,3.007969617843628)(487,3.6428873538970947)(488,3.770678997039795)(489,3.015061140060425)(490,2.256117105484009)(491,3.3204293251037598)(492,4.200146675109863)(493,3.5890586376190186)(494,3.461545944213867)(495,3.2048628330230713)(496,3.7057621479034424)(497,3.565397262573242)(498,3.7431371212005615)(499,3.1653826236724854)(500,4.036604404449463)(501,4.8336873054504395)(502,4.54766845703125)(503,4.07742977142334)(504,4.29622220993042)(505,4.063142776489258)(506,4.971269130706787)(507,3.613506317138672)(508,3.6543736457824707)(509,3.495379686355591)(510,3.859546661376953)(511,3.5067362785339355)(512,3.428370952606201)(513,2.618124485015869)(514,4.80186653137207)(515,1.4693450927734375)(516,2.13703989982605)(517,3.4423975944519043)(518,2.529195547103882)(519,1.8513039350509644)(520,2.82438588142395)(521,2.813863515853882)(522,1.9515769481658936)(523,2.0890631675720215)(524,2.9847612380981445)(525,2.6367990970611572)(526,2.557589530944824)(527,2.095872640609741)(528,2.1181867122650146)(529,3.291567087173462)(530,1.7510144710540771)(531,4.229649543762207)(532,1.5325626134872437)(533,4.044524192810059)(534,3.0318918228149414)(535,1.7269134521484375)(536,2.450186014175415)(537,1.8106049299240112)(538,1.7903510332107544)(539,4.683688640594482)(540,1.6008728742599487)(541,1.3112034797668457)(542,4.643749237060547)(543,4.132994651794434)(544,2.094921588897705)(545,2.971599817276001)(546,4.078929901123047)(547,0.7739177942276001)(548,2.5216128826141357)(549,2.8121354579925537)(550,0.5964282155036926)(551,3.5797204971313477)(552,2.7993409633636475)(553,1.9649006128311157)(554,2.983957529067993)(555,3.6980528831481934)(556,1.6200063228607178)(557,1.333514928817749)(558,2.580685615539551)(559,1.9900879859924316)(560,2.1398680210113525)(561,1.938896894454956)(562,1.152321219444275)(563,3.7988059520721436)(564,1.1819878816604614)(565,1.3584520816802979)(566,1.8167407512664795)(567,2.784320116043091)(568,3.768495559692383)(569,0.9201547503471375)(570,4.310686111450195)(571,2.6670784950256348)(572,2.192180633544922)(573,2.6648848056793213)(574,4.149143695831299)(575,3.4088151454925537)(576,2.8839242458343506)(577,1.4168641567230225)(578,0.7884765863418579)(579,4.078624725341797)(580,1.17002272605896)(581,1.88533616065979)(582,2.0779969692230225)(583,1.485579013824463)(584,1.749842882156372)(585,2.8120603561401367)(586,3.798488140106201)(587,2.4498989582061768)(588,2.488903045654297)(589,3.101536273956299)(590,2.550597667694092)(591,3.546492099761963)(592,0.4210253357887268)(593,2.9009134769439697)(594,3.042717695236206)(595,2.7946536540985107)(596,1.8623123168945312)(597,3.667670249938965)(598,4.506235599517822)(599,1.0289708375930786)(600,3.82393217086792)(601,1.7996925115585327)(602,1.2397103309631348)(603,4.16265344619751)(604,2.097740650177002)(605,3.1479854583740234)(606,2.7577133178710938)(607,3.079536199569702)(608,3.2765347957611084)(609,4.281512260437012)(610,3.2310872077941895)(611,2.288017511367798)(612,2.120605230331421)(613,2.4441826343536377)(614,2.4741432666778564)(615,4.05692195892334)(616,4.5012736320495605)(617,1.0912569761276245)(618,4.8333940505981445)(619,0.7084862589836121)(620,3.3615963459014893)(621,4.3586015701293945)(622,0.363944947719574)(623,2.336825370788574)(624,2.6289494037628174)(625,0.864477813243866)(626,1.9454299211502075)(627,2.18815541267395)(628,1.5624709129333496)(629,1.9528371095657349)(630,3.643763303756714)(631,1.7950842380523682)(632,2.4605815410614014)(633,1.2974822521209717)(634,1.8650591373443604)(635,3.6441915035247803)(636,2.1391730308532715)(637,1.8494758605957031)(638,3.024895429611206)(639,2.1483635902404785)(640,1.9137529134750366)(641,1.8615800142288208)(642,3.1778836250305176)(643,1.3386669158935547)(644,2.823662757873535)(645,2.5742154121398926)(646,1.8481085300445557)(647,3.471597194671631)(648,3.7057840824127197)(649,3.3955278396606445)(650,3.74544620513916)(651,4.021198272705078)(652,2.6713345050811768)(653,2.7367849349975586)(654,2.5299718379974365)(655,3.0944621562957764)(656,3.199086904525757)(657,2.3045413494110107)(658,2.5273027420043945)(659,5.195677280426025)(660,0.792182207107544)(661,4.356198310852051)(662,1.4417263269424438)(663,1.9637008905410767)(664,2.5842020511627197)(665,3.026578426361084)(666,1.572898030281067)(667,3.2756881713867188)(668,3.095790147781372)(669,3.6162333488464355)(670,4.4702277183532715)(671,0.6502029895782471)(672,2.9084582328796387)(673,3.7756879329681396)(674,0.9936001300811768)(675,3.337287425994873)(676,1.6274657249450684)(677,3.2290632724761963)(678,3.352159023284912)(679,0.8220257759094238)(680,2.1172187328338623)(681,4.060834884643555)(682,1.8706855773925781)(683,1.2804316282272339)(684,2.4138941764831543)(685,5.148870468139648)(686,1.9189614057540894)(687,1.540899634361267)(688,2.465848684310913)(689,3.356243848800659)(690,2.891585111618042)(691,2.5304009914398193)(692,3.578798770904541)(693,3.057770013809204)(694,3.494809865951538)(695,7.827620506286621)(696,2.2233588695526123)(697,1.9403051137924194)(698,3.495450019836426)(699,3.3474929332733154)(700,5.734088897705078)(701,2.7045648097991943)(702,2.9012794494628906)(703,2.5417966842651367)(704,3.4946370124816895)(705,2.270542621612549)(706,3.103794813156128)(707,2.803892135620117)(708,2.1809592247009277)(709,3.8736307621002197)(710,1.5389931201934814)(711,3.642505645751953)(712,1.3562649488449097)(713,2.341174602508545)(714,2.603156089782715)(715,3.450073003768921)(716,2.225855588912964)(717,4.387012958526611)(718,4.847173690795898)(719,1.4644118547439575)(720,2.841940402984619)(721,3.3268792629241943)(722,2.419736385345459)(723,1.402251124382019)(724,2.728292465209961)(725,2.016329765319824)(726,2.081200361251831)(727,1.1972545385360718)(728,1.895851492881775)(729,4.316678524017334)(730,4.89585542678833)(731,1.5680932998657227)(732,1.731312870979309)(733,2.6550700664520264)(734,1.910584807395935)(735,1.9025338888168335)(736,3.3639931678771973)(737,3.7417893409729004)(738,5.456025123596191)(739,3.311128854751587)(740,1.8924167156219482)(741,6.260453224182129)(742,5.249344825744629)(743,4.338598251342773)(744,4.1908955574035645)(745,3.8671422004699707)(746,4.642164707183838)(747,8.31470775604248)(748,1.414919376373291)(749,2.432551622390747)(750,2.7719058990478516)(751,2.796525239944458)(752,2.036053419113159)(753,4.053290367126465)(754,2.7060582637786865)(755,1.8930240869522095)(756,2.775203227996826)(757,5.496347427368164)(758,1.3446611166000366)(759,1.968492031097412)(760,4.281970024108887)(761,1.7599254846572876)(762,1.3340238332748413)(763,3.0217723846435547)(764,2.773610830307007)(765,2.4429521560668945)(766,2.4170079231262207)(767,2.0558834075927734)(768,4.3603949546813965)(769,5.505387783050537)(770,3.991619825363159)(771,3.6032416820526123)(772,3.2042043209075928)(773,3.0507729053497314)(774,3.263658285140991)(775,5.50120735168457)(776,3.9128286838531494)(777,2.470282793045044)(778,2.562939167022705)(779,3.247553586959839)(780,5.245779514312744)(781,4.602224349975586)(782,4.314232349395752)(783,4.232008457183838)(784,3.7426726818084717)(785,4.619677543640137)(786,3.282815933227539)(787,2.6284375190734863)(788,3.9908037185668945)(789,5.466851234436035)(790,4.686559200286865)(791,3.799253463745117)(792,3.80330753326416)(793,5.203044891357422)(794,0.9597439765930176)(795,1.0645432472229004)(796,1.9269697666168213)(797,3.006695508956909)(798,5.082086563110352)(799,1.6025066375732422)(800,1.9011268615722656)(801,2.9527227878570557)(802,3.274630069732666)(803,1.7150899171829224)(804,4.094541549682617)(805,2.6681153774261475)(806,1.6214120388031006)(807,3.8119726181030273)(808,2.8155083656311035)(809,2.0504281520843506)(810,1.8014817237854004)(811,3.2369046211242676)(812,5.697021007537842)(813,3.9831972122192383)(814,2.49924373626709)(815,4.022495269775391)(816,3.7245125770568848)(817,2.834684133529663)(818,2.7357699871063232)(819,2.3449323177337646)(820,3.2100682258605957)(821,2.979681968688965)(822,3.8067879676818848)(823,3.5193932056427)(824,4.575345516204834)(825,1.0882834196090698)(826,1.4832755327224731)(827,2.972700834274292)(828,2.6548080444335938)(829,3.0368831157684326)(830,2.4846842288970947)(831,3.443697929382324)(832,4.784514427185059)(833,2.665996551513672)(834,2.5285415649414062)(835,2.852404832839966)(836,4.811949729919434)(837,2.6885173320770264)(838,3.07306170463562)(839,1.8609628677368164)(840,2.4631786346435547)(841,3.273930549621582)(842,1.754103422164917)(843,2.7296364307403564)(844,4.9254865646362305)(845,3.861628770828247)(846,2.87898588180542)(847,2.4379589557647705)(848,2.828974485397339)(849,2.3313090801239014)(850,2.108321189880371)(851,4.462730407714844)(852,5.670910358428955)(853,1.2361781597137451)(854,2.786106824874878)(855,2.365065097808838)(856,2.3747971057891846)(857,2.5397613048553467)(858,2.5805389881134033)(859,4.277507305145264)(860,4.066898822784424)(861,3.5625112056732178)(862,4.224390506744385)(863,4.506627559661865)(864,6.507246017456055)(865,2.8592307567596436)(866,2.9105587005615234)(867,3.0709664821624756)(868,3.242190361022949)(869,2.5423836708068848)(870,4.927700996398926)(871,4.218799591064453)(872,3.546276807785034)(873,5.2118659019470215)(874,8.505857467651367)(875,2.0112740993499756)(876,3.0503134727478027)(877,3.113710641860962)(878,5.6910834312438965)(879,6.479055881500244)(880,0.8903230428695679)(881,2.037933111190796)(882,4.352322101593018)(883,3.0902953147888184)(884,1.6453063488006592)(885,3.3183891773223877)(886,2.894347667694092)(887,5.243115425109863)(888,2.3196818828582764)(889,2.80594801902771)(890,3.41117787361145)(891,2.9241652488708496)(892,1.6119651794433594)(893,2.4286813735961914)(894,3.2464098930358887)(895,7.211609363555908)(896,1.8093911409378052)(897,2.4987926483154297)(898,3.818460702896118)(899,4.268970489501953)(900,1.548043131828308)(901,2.4635672569274902)(902,3.13063645362854)(903,6.168071746826172)(904,3.2073147296905518)(905,3.711583137512207)(906,4.2263360023498535)(907,3.264611005783081)(908,5.694896697998047)(909,8.81506633758545)(910,4.296365261077881)(911,2.555182933807373)(912,2.4954311847686768)(913,4.180715084075928)(914,4.543162822723389)(915,2.8261544704437256)(916,1.8807544708251953)(917,4.141933917999268)(918,3.1226601600646973)(919,2.4501569271087646)(920,3.369523048400879)(921,3.6237311363220215)(922,2.6187288761138916)(923,3.522218704223633)(924,6.345345497131348)(925,1.8559719324111938)(926,2.5754151344299316)(927,2.9096808433532715)(928,3.9123435020446777)(929,2.236919641494751)(930,2.4029505252838135)(931,4.803875923156738)(932,3.4647769927978516)(933,5.3098039627075195)(934,1.9197742938995361)(935,2.377455949783325)(936,3.135953187942505)(937,2.6468489170074463)(938,2.005323648452759)(939,2.3207900524139404)(940,3.2150259017944336)(941,2.4651622772216797)(942,3.1980557441711426)(943,3.747084617614746)(944,7.065956115722656)(945,1.6180742979049683)(946,2.134354591369629)(947,3.244738817214966)(948,1.4495419263839722)(949,1.183156132698059)(950,2.2630958557128906)(951,2.0426037311553955)(952,3.3171353340148926)(953,5.310707092285156)(954,1.7237075567245483)(955,2.27325701713562)(956,2.6634414196014404)(957,2.6261544227600098)(958,5.851193428039551)(959,2.421734094619751)(960,2.4557130336761475)(961,2.770468235015869)(962,3.6418540477752686)(963,6.163984298706055)(964,2.762871265411377)(965,1.6466434001922607)(966,3.0605101585388184)(967,4.246766090393066)(968,2.6375668048858643)(969,1.4553325176239014)(970,2.307533025741577)(971,2.2435083389282227)(972,3.0652732849121094)(973,3.276211977005005)(974,2.0970935821533203)(975,2.1779134273529053)(976,2.439866781234741)(977,3.4533281326293945)(978,2.279244899749756)(979,3.2979235649108887)(980,1.8792721033096313)(981,2.7703826427459717)(982,2.452730894088745)(983,3.046931505203247)(984,3.8639612197875977)(985,1.7796053886413574)(986,2.9201853275299072)(987,4.781557559967041)(988,2.4456522464752197)(989,2.7837328910827637)(990,2.6842470169067383)(991,3.4791886806488037)(992,2.999601364135742)(993,4.6655778884887695)(994,3.53776216506958)(995,3.158442497253418)(996,3.06606125831604)(997,4.922928810119629)(998,3.638472080230713)(999,1.7781627178192139)(1000,4.209768772125244)

    };
    \legend{DCGAN, Age-ACGAN}
 
\end{axis}
\end{tikzpicture}
\caption{Comparison of generator convergence speed between DCGAN and our proposed network Age-ACGAN. Generator loss of Age-ACGAN converges after 500 iterations, while DCGAN experiences constant increase in generator loss in the first 1000 iterations.}
\end{figure}
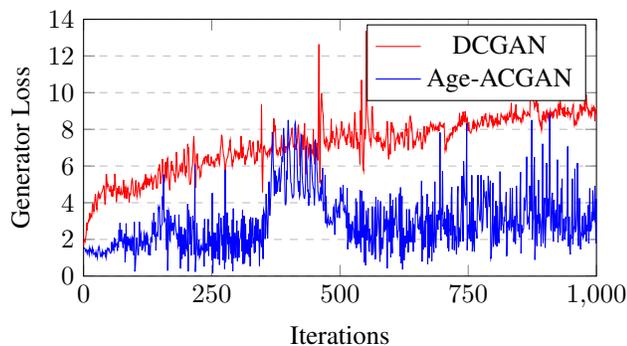

\subsection{Age Conditioned Synthetic Images}
\label{ssec:Age Conditioned Synthetic Images}
The synthetic CT images along with their pancreas segmentation masks for each age class were shown in Fig. 4. In the synthesized image, we were able to observe a clear trend in which the pancreas becomes gradually elongated as the patient age increases. This shows Age-ACGAN's capability of capturing realistic growth trends in the pancreas.
\begin{figure}[h]
\begin{minipage}[b]{1.0\linewidth}
 \centerline{\textbf{Infant}~~~~~~~~~~~~~~~\textbf{Preschool}~~~~~~~~~~~~~\textbf{Adolescent}}\medskip
 
 \centerline{\includegraphics[width=8.5cm, height=1.5cm]{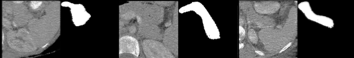}}
\end{minipage}
\caption{CT with segmentation mask synthesis results. Test images were conditionally synthesized with a vector denoting the desired age classes. The synthesized pancreas masks is observed to become elongated as the patient age increases. This demonstrates our network's ability to capture growth trends in the data distribution.}
\end{figure}

\section{Conclusions}
\label{sec:results}

In this study, we propose an age-conditioned generative adversarial network for the synthesis of CT images and the corresponding segmentation mask. From the results of our preliminary experiments, our proposed network is capable of synthesizing high-resolution patches from abdominal CT images along with their pancreas segmentation masks. It is also able to capture realistic trends in pancreas growth during age-conditioned synthesis. Our work can be further extended to jointly synthesize CT images and segmentation masks for other organs, \color{black} and shape arithmetic can be performed to combine learned latent spaces to create new image-mask pairs \color{black}.

\bibliographystyle{IEEEbib}
\bibliography{refs}

\end{document}